# Observation of ultraslow hole dynamics in the 3D topological insulator $Bi_2Se_3$ coated with a thin $MgF_2$ layer using multiphoton pumped UV-Vis transient absorption spectroscopy

Yuri D. Glinka[1,2, a)], Tingchao He[3, a)], Xiao Wei Sun[1,4, a)]

[1] *Guangdong University Key Lab for Advanced Quantum Dot Displays and Lighting, Shenzhen Key Laboratory for Advanced Quantum Dot Displays and Lighting, Department of Electrical and Electronic Engineering, Southern University of Science and Technology, Shenzhen 518055, China*
[2] *Institute of Physics, National Academy of Sciences of Ukraine, Kyiv 03028, Ukraine*
[3] *College of Physics and Energy, Shenzhen University, Shenzhen 518060, China*
[4] *Shenzhen Planck Innovation Technologies Pte Ltd., Longgang, Shenzhen 518112, China*

[a)] Author to whom correspondence should be addressed. Electronic mail: yuridglinka@yahoo.com, tche@szu.edu.cn, sunxw@sustech.edu.cn

Individual relaxation dynamics of electrons and holes in optically pumped semiconductors is rarely observed due to their overlap. Here we report the individual dynamics of long-lived (~200 μs) holes observed at room temperature in a 10 nm thick film of the 3D topological insulator (TI) $Bi_2Se_3$ coated with a 10 nm thick $MgF_2$ layer using transient absorption spectroscopy in the UV-Vis region. The ultraslow hole dynamics was observed by applying multiphoton resonant pumping of massless Dirac fermions and bound valence electrons in $Bi_2Se_3$ at a certain wavelength sufficient for their photoemission and subsequent trapping at the $Bi_2Se_3/MgF_2$ interface. The emerging deficit of electrons in the film makes it impossible for the remaining holes to recombine, thus causing their ultraslow dynamics measured at a specific probing wavelength. We also found an extremely long rise time (~600 ps) for this ultraslow optical response, which is due to the large spin-orbit coupling (SOC) splitting at the valence band maximum and the resulting intervalley scattering between the splitting components. The ultraslow hole dynamics in $Bi_2Se_3$ due to the presence of the $Bi_2Se_3/MgF_2$ interface is nevertheless much faster than the known ultraslow electron dynamics at the $Si/SiO_2$ interface, also induced by multiphoton excitation in Si. The observed dynamics of long-lived holes is gradually suppressed with decreasing $Bi_2Se_3$ film thickness for the 2D TI $Bi_2Se_3$ (film thickness 5, 4 and 2 nm) due to the loss of resonance conditions for multiphoton photoemission caused by the gap opening at the Dirac surface state nodes. This behavior indicates that the dynamics of massive Dirac fermions predominantly determines the relaxation of photoexcited carriers for both the 2D topologically nontrivial and 2D topologically trivial insulator phases.

## I. INTRODUCTION

Optical pump-probe techniques are widely used to study ultrafast carrier dynamics in semiconductors due to their simplicity and the wide variety of ultrashort pulses with specific photon energies provided by commercially available ultrafast lasers.[1-12] However, the study of the individual relaxation dynamics of electrons and holes in optically pumped semiconductors, despite the fact that they involve the same processes associated with electron/hole inelastic scattering by longitudinal optical (LO) and acoustic phonons [Fig. 1(a)],[8-12] remains a difficult problem. The reason is that both dynamics simultaneously contribute to the modulation of the complex refractive index of the material and, therefore, affect the corresponding optical probing.[1-14] To avoid this limitation, special conditions are required. In particular, this problem can be partially solved by using doped semiconductors with a carrier concentration of $10^{16}$-$10^{18}$ cm$^{-3}$. However, the optical pumping in this case must be extremely weak in order to generate a carrier density well below the doping level, which makes probing difficult.[7] Another option is the spatial separation of electrons and holes, which can be achieved in some semiconductor heterostructures and topological insulators (TIs).[11,12,15-19] However, since the penetration depth of pumping and probing light usually exceeds the effective width of the heterojunction in heterostructures and Dirac surface states (SS) in TIs, the relaxation dynamics of electrons is superimposed on the dynamics of holes. Nevertheless, the electron dynamics can be recognized, since holes usually relax faster [Fig. 1(a)].[8-12] Similarly, in materials exhibiting large spin-orbit coupling (SOC) splitting at the valence band (VB) maximum (for example, transition metal dichalcogenides), the individual relaxation dynamics of holes can also be observed. In the latter case, this recognition is possible because the hole dynamics is much slower than the electron dynamics and persists beyond a few nanoseconds, since the hole relaxation becomes equivalent to the intervalley relaxation.[20-22] It should be noted that the SOC splitting in the TI $Bi_2Se_3$ is large enough to expect intervalley relaxation of holes as well.[23,24] In general, the individual relaxation dynamics of electrons and holes in optically pumped semiconductors is rarely observed, and its assignment is usually controversial.

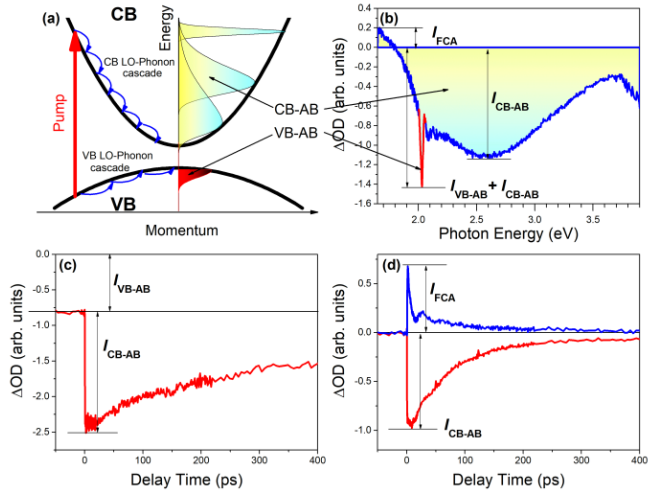

FIG. 1. (a) Schematic representation of the band structure of a semiconductor. The one-photon pump transition (red vertical up arrow) originates from the valence band (VB) states and ends in the conduction band (CB) states. The corresponding LO-phonon cascades and the assumed transient electron/hole distributions during relaxation are shown. (b), (c), and (d) Example of TA spectra of the TI $Bi_2Se_3$ measured at 2.07 eV (600 nm) pumping and the typical pump-probe traces. The intensities of the background signal associated with VB absorption bleaching (VB-AB), as well as the CB absorption bleaching (CB-AB) and the inverse bremsstrahlung type free carrier absorption (FCA) signals are indicated.

To illustrate controversial assignments, we consider the negative background signal from $Bi_2Se_3$ thin films coated with a 10 nm thick $MgF_2$ layer, which was observed in pump-probe reflectivity experiments using ultrashort (~100 fs) pump and probe pulses of the same wavelength (820 nm - photon energy ~1.51 eV) with a repetition rate of 80 MHz (time interval between successive pulses ~12.5 ns).[11,19] This signal is additional to the usual pump-probe signal, but does not depend on the delay time between the pump and probe pulses (Fig. 1). The discussed background signal has also been observed in pump-probe second harmonic generation, but with a positive amplitude.[19] Both background signals show the same type of dependence on pump power as the usual time-dependent pump-probe signal (linear and quadratic, respectively).[11,19] Also, they are reset to zero when either the pump or probe beam is blocked, so they have been assigned to the pump-probe signals. To explain this unusual behavior, it was assumed that the background signals are determined by long-lived (slowly decaying) carriers with a decay-time constant comparable to the reciprocal repetition rate of laser pulses. This assumption means that between two successive laser pulses the relaxation of carriers does not occur completely, thus leading to their progressive accumulation as long as the sample is irradiated with laser light. Accordingly, the background signals were initially attributed to the quasi-stationary electron localization in Dirac SS on opposite film surfaces,[11] and to a resulting capacitor-like electric field.[19] It is important that the contribution of long-lived carriers to the relaxation dynamics (to the corresponding intensity of background signals) gradually decreases with decreasing film thickness.[25] This behavior was explained by an increase in the rate of recombination of massless Dirac fermions in the higher energy Dirac SS (SS2) and holes in the lowest energy Dirac SS (SS1).

Subsequently, employing transient absorption (TA) spectroscopy in the UV-Vis region [Fig. 1(b)], we found that the negative background signal from the TI $Bi_2Se_3$ appears in the pump-probe traces only when using the same pump and probe wavelengths. As a result, the signal was assigned to the ultraslow dynamics of holes at the VB top and the resulting narrowband absorption bleaching (VB-AB) [Fig. 1(a)-(c)].[26-29] Absorption bleaching (AB) means the filling of the phase-space in Dirac SS and bulk states, followed by their Pauli blocking (the generalized Burstein-Moss effect).[28,30] In contrast, much faster electron relaxation dynamics in the conduction band (CB) and the resulting broadband absorption bleaching (CB-AB) appear when probing at a wavelength other than pumping and persist beyond a few hundred picoseconds (Fig. 1). The latter behavior is hence in good agreement with what has been observed in numerous optical pump-probe experiments,[11,12,19,25-29,31-33] as well as in experiments using optical pumping and other non-optical methods of probing.[34-44]

Surprisingly, the background VB-AB signal from the TI $Bi_2Se_3$ has been observed using a laser with a pulse repetition rate of 1 kHz [time interval between successive ultrashort (~100 fs) pulses ~1 ms].[26-29] This behavior means that the decay time of the long-lived holes should be comparable to ~1 ms. Such a slow process, even though it does not affect the usual pump-probe signal, has never been observed for the TI $Bi_2Se_3$. Since this relaxation dynamics is too slow to be attributed to intervalley relaxation of holes, some additional process must be considered responsible for this ultraslow behavior. In addition, free carrier absorption (FCA) further confuses the relaxation dynamics of electrons and holes with a positive contribution in TA spectra [Fig. 1(b) and (d)]. This contribution strongly depends on the density of photoexcited carriers and the mechanisms of their interaction with the crystalline lattice (inelastic/quasi-elastic) and, therefore, manifests itself through either Drude absorption[30,45] or inverse bremsstrahlung type absorption.[26-30]

Here we report the observation of the individual relaxation dynamics of long-lived (~200 µs) holes in the 3D TI $Bi_2Se_3$ [10 quintuple layers (QL) thick film coated with a 10 nm thick $MgF_2$ layer] at room temperature using TA spectroscopy in the UV-Vis region, where QL represents five covalently bonded Se–Bi–Se–Bi–Se atomic sheets with a total thickness of ~1 nm. All measurements were performed under pumping by ultrashort (~100 fs) pulses with a pulse repetition rate of 1 kHz at different wavelengths of 340, 400, 500, 600, and 730 nm (the corresponding photon energies of ~3.65, 3.1, 2.48, 2.07, and 1.7 eV, respectively) and probing in the UV-Vis region (1.65 - 3.8 eV).

Ultraslow hole dynamics was observed by applying multiphoton resonant pumping of massless Dirac fermions and bound valence electrons in $Bi_2Se_3$ at a certain wavelength sufficient for their photoemission and subsequent trapping at the $Bi_2Se_3/MgF_2$ interface. The resulting loss of electrons in the film makes it impossible for the remaining holes to recombine and creates their long-lived excess, which is detected at a specific probe wavelength as an ultraslowly decaying response characterizing the electron detrapping rate. Correspondingly, the effect sharply decreases with decreasing pump photon energy, since a higher order nonlinearity (higher order multiphoton process) is required to exceed the $Bi_2Se_3$ work function of 5.97 eV.[28,35] We also found an extremely long rise time (~600 ps) for this ultraslow optical response, which is explained by the large SOC splitting at the VB maximum and the resulting intervalley relaxation between the splitting components.

The dynamics of long-lived holes is progressively suppressed with decreasing film thickness for 2D TI $Bi_2Se_3$ (film thicknesses below 6 QL,[45]). This behavior unambiguously indicates that in the 2D TI $Bi_2Se_3$, multiphoton photoemission of massless Dirac fermions becomes inefficient due to the loss of resonance conditions caused by the gap opening at the Dirac SS nodes.[46] As a result, massive Dirac fermions[46-51] predominantly determines the relaxation of photoexcited carriers for both the 2D topologically nontrivial and 2D topologically trivial insulator phases. The observed ultraslow hole dynamics in $Bi_2Se_3$ due to the presence of the $Bi_2Se_3/MgF_2$ interface is nevertheless much faster than the known ultraslow electron dynamics at the $Si/SiO_2$ interface, also induced by multiphoton excitation in Si.[52,53] Finally, we concluded that the observed dynamics of long-lived holes in the 3D TI $Bi_2Se_3$ coated with a thin $MgF_2$ layer can be applied to create new thin-film quantum materials that are promising for modern optoelectronics and optospintronics.

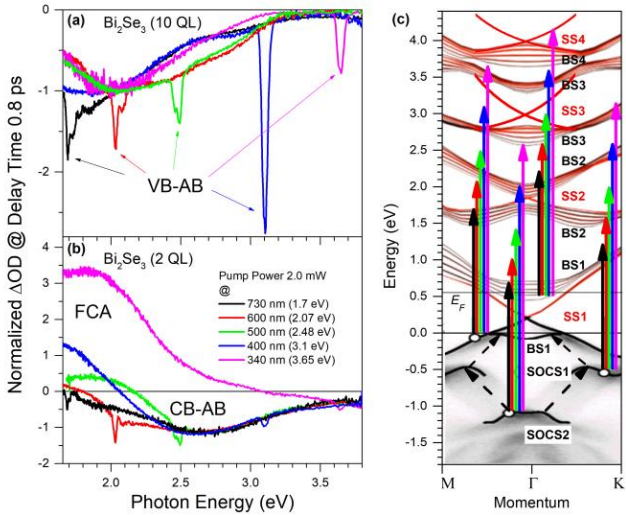

FIG. 2. (a) and (b) TA spectra of the 3D TI $Bi_2Se_3$ (film thickness 10 QL) and the 2D TI $Bi_2Se_3$ (film thickness 2 QL), respectively, measured at delay time of ~0.8 ps using pump pulses of certain wavelengths indicated by the corresponding colors. The pump power for all wavelengths was ~2 mW. Narrow peaks associated with the VB-AB contribution and broadband CB-AB and FCA contributions are shown. (c) The band structure of the 3D TI $Bi_2Se_3$ calculated in Ref. 23 and 35. Zero energy has been shifted to the top of the VB. The bulk states and Dirac surface states are marked as BS and SS, respectively. The higher energy Dirac SS (SS3 and SS4) predicted in Ref. 27 are shown. The one-photon pump transitions (color vertical up arrows) originating from the VB top and the VB SOC split states (SOCS1 and SOCS2) of the bulk, as well as from the Dirac surface states (SS1) below the Fermi energy ($E_F$) are shown. The color of the arrows and their length correspond to the energies of the pump photons, as indicated in (b). Dashed arrows show intervalley processes of hole relaxation.

## II. EXPERIMENTAL DETAILS

### A. Sample preparation.

The 2, 4, 5, and 10 nm thick $Bi_2Se_3$ films were grown on 0.5 mm $Al_2O_3(0001)$ substrates by molecular beam epitaxy, with a 10 nm thick $MgF_2$ protecting capping layer, which was grown at room temperature without exposing the film to the atmosphere. The films have been found to be epitaxial and the nominal number of QL was accurate to approximately 5%. The level of disorder of the grown films and their quality were approximately the same.[54] Furthermore, the films reveal the *n*-type doping measured using the Hall effect.[55] The doping level increases with decreasing film thickness from ~$2.0\times10^{19}$ to ~$5.0\times10^{19}$ $cm^{-3}$.

### B. Experimental setup.

TA spectra were measured using a TA spectrometer (Newport), which was equipped with a Spectra-Physics Solstice Ace regenerative amplifier (~100 fs pulses at 800 nm with 1.0 KHz repetition rate) to generate the supercontinuum probing beam and a Topas light convertor for the pumping beam. The spectrometer has also been modified to suppress all coherent artifacts emanating from the sapphire substrates and appearing on a subpicosecond time scale, as discussed elsewhere.[26-30]

We used optical pumping at 340, 400, 500, 600, and 730 nm (the corresponding photon energies of ~3.65, 3.1, 2.48, 2.07, and 1.7 eV, respectively) and a supercontinuum probing beam generated in a sapphire plate and spread across the UV-Vis spectral region from 326 to 751 nm (~1.65 to ~3.8 eV). The probing beam was at normal incidence, while the pumping beam was at an incident angle of ~30°. All measurements were performed in air and at room temperature using a cross-linear-polarized geometry. Specifically, the pumping and probing beams were polarized out-of-plane (vertical) and in-plane (horizontal) of incidence, respectively. The data matrix was corrected for the chirp of the supercontinuum probing pulse using a coherent nonlinear optical response (presumably degenerate four-wave mixing or two-photon absorption) from a thin (0.3 mm) sapphire plate.[28,29,56,57]

The spot sizes of the pumping and probing beams were ~400 and ~150 μm, respectively. The average pump beam power could vary within the range of 0.5 - 2.0 mW, which corresponds to the pumping intensity (power density) of 4.0 - 16.0 GW $cm^{-2}$. The broadband probing beam was of ~0.4 mW average power, which for the same as the pumping beam bandwidth (~26 meV) provides the probing intensity of ~0.15 GW $cm^{-2}$. Since the latter value is much smaller than that of the pump beam, the influence of the probing beam on carrier excitation was negligible.

## III. RESULTS AND DISCUSSION

Figure 2(a) and (b) show the TA spectra of a 10 QL thick $Bi_2Se_3$ film (3D TI) and a 2 QL thick $Bi_2Se_3$ film (2D TI) coated with a 10 nm thick $MgF_2$ layer, which were measured using pumping at different wavelengths in the range from 340 to 730 nm (photon energies from 3.65 to 1.7 eV, respectively) with an average pump power of ~2.0 mW. The broadband contribution, extending from ~1.65 to ~3.8 eV, develops on a subpicosecond time scale, which is in full agreement with our previous measurements with pumping by photons with energies of 3.65 and 1.7 eV of various powers.[26-29] The measured TA spectra of the 2D TI $Bi_2Se_3$ consists of negative and positive contributions, while only negative contributions appear for the 3D TI $Bi_2Se_3$. The positive contribution is due to the inverse bremsstrahlung type FCA in the Dirac SS2, while the negative contributions are associated with the CB-AB (a broadband negative contribution) and the VB-AB (a narrowband negative contribution peaked near the pumping photon energies) [Figs. 1(b) and 2(a), 2(b)]. It is worth noting here that the effect of scattered pump light in the detecting branch of the setup was completely suppressed by the cross-linearly polarized configuration used in our experiments, and therefore it should not be confused with the observed ultraslow VB-AB contribution.

We also note that the assignment of the bands in TA spectra differs from that commonly used for ultrafast TA spectroscopy of molecules, which typically deals with ground-state bleaching, stimulated emission, excited-state absorption, and product absorption.[58] Obviously, this terminology referring to localized molecular orbitals cannot be applied to semiconductor dealing with delocalized free carriers in the CB/VB and taking into account their Fermi-Dirac distribution and the electron-phonon (hole-phonon) interaction. Since confusion in terminology is common in the literature, we indicate here several basic principles of TA spectroscopy of semiconductors. Correspondingly, AB is determined by the Fermi-Dirac occupancy factors for electrons and holes in the CB and VB, respectively.[4] Cooling of carriers due to inelastic scattering by LO phonons (Frohlich relaxation mechanism) is extremely fast ($\leq 3$ ps) and weakly depends on the photoexcited carrier density.[11] In addition, free (non-interacting) electrons (holes) in the CB (VB) cannot absorb light due to the energy-momentum conservation restrictions.[30,59] However, there are two exceptions that allow the absorption of light by free carriers associated with their collective excitation (the spectral region below the plasma frequency - Drude type FCA),[30,45,59] and the absorption of light by free carriers when they collide with the lattice atoms (the spectral region above the plasma frequency - the inverse bremsstrahlung type FCA).[26-30] It would seem that the application of the molecular approach to characterize the features of the TA spectra of semiconductors seems possible in the presence of localization of carriers in certain states of the CB (VB). For example, in the 3D TI $Bi_2Se_3$, photoexcited electrons are accumulated in the Dirac SS2 before recombination. However, the positive contribution to the TA spectra [Fig. 2(b)] cannot be associated with the excited-state absorption, as is usually the case in the TA spectra of molecules. The reason is that the spectrum is strongly extended towards higher energies far from the Dirac SS2, has a smooth, Gaussian-like profile, and does not follow any features of the band structure of $Bi_2Se_3$. All these facts indicate that in the Dirac SS2, the inverse bremsstrahlung type FCA is realized, which does not require the participation of any higher energy states of $Bi_2Se_3$, as might be expected in the molecular approach.

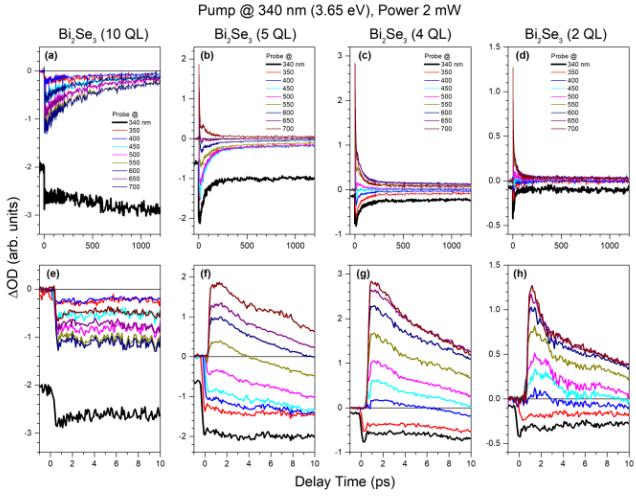

FIG. 3. Pump-probe traces of the 3D TI $Bi_2Se_3$ (10 QL thick film) and the 2D TI $Bi_2Se_3$ (5, 4, and 2 QL thick films) measured at different probing wavelengths, as indicated by the corresponding colors, using the 340 nm pumping (3.65 eV photon energy) of ~2.0 mW average power. For clarity, the bottom row plots show the same set of pump-probe traces as the top row plots, but on a shorter delay time scale.

It is clearly seen that the VB-AB contributions for the 3D TI $Bi_2Se_3$ are much stronger than those for the 2D TI $Bi_2Se_3$ [Fig. 2(a) and (b)]. In addition, the relative intensity of the VB-AB contributions for the 3D TI $Bi_2Se_3$ depends nonmonotonically on the pump photon energy. This behavior manifests itself as resonance features (as seen well for 400 nm pumping). On the contrary, the relative intensity of the much weaker VB-AB contributions for the 2D TI $Bi_2Se_3$ increases rather monotonically with decreasing pump photon energy. Finally, for the 2D TI $Bi_2Se_3$, the inverse bremsstrahlung type FCA gradually increases with increasing pump photon energy.

All these observations point to a complex behavior of photoexcited carriers in the 2D TI $Bi_2Se_3$, which differs significantly from that in the 3D TI $Bi_2Se_3$. First, we only note that the bulk effect decreases with decreasing film thickness, and the dynamics of Dirac fermions in the 2D TI $Bi_2Se_3$ determines predominantly the relaxation of photoexcited carriers.[26-29] Correspondingly, the shape of the broadband CB-AB contribution for the 3D TI $Bi_2Se_3$ closely matches the imaginary part of the dielectric function of $Bi_2Se_3$ and hence its CB density of states.[60] For the 2D TI $Bi_2Se_3$, the CB-AB contribution gradually decreases with decreasing film thickness, and its maximum shifts towards higher energies due to higher energy bulk states.[28] The FCA contribution tends to a maximum in the visible region, which energetically corresponds to the Dirac SS2, although at concentrations of photoexcited carriers in the film ($10^{19}$ - $10^{20}$ cm$^{-3}$) the plasma frequency is in the IR region.[30,45] Consequently, the Drude mechanism of FCA in the visible region is ineffective and the inverse bremsstrahlung type FCA in the Dirac SS2 dominates.[26]

The discussed trends are more clearly seen in the corresponding pump-probe traces (Figs. 3 and 4). In general, the relaxation dynamics manifesting in negative-amplitude pump-probe traces are due to ultrafast decay associated with Frohlich relaxation mechanism, followed by a longer decay associated with the excitation of acoustic phonons/plasmons and, finally, with electron-hole recombination.[10-12,19,25-29,31-44] The observed ultrafast relaxation reveal hence all features typical of the TI $Bi_2Se_3$. In particular, the rise time of pump-probe traces is determined by the electron-electron (hole-hole) thermalization and is characterized by a rise time constant of 0.15 ps.[25-29] Further relaxation for the 3D TI $Bi_2Se_3$ is associated with the emission of a cascade of LO phonons for several ps, which is accompanied by a spatial redistribution of electrons and subsequent excitation of acoustic phonons for ~10 ps.[10-12,19,26,28] The final stage of relaxation is electron-hole recombination for several hundreds of ps.[25-29] For the 2D TI $Bi_2Se_3$, the additional quasi-elastic scattering of Dirac fermions by surface LO phonons contributes to positive-amplitude pump-probe traces characterizing the inverse bremsstrahlung type FCA in the Dirac SS2.[26-29] Further relaxation in the Dirac SS2 is due to the excitation of acoustic plasmons, which is also followed by electron-hole recombination.[19,26,27]

However, the most important feature that we pay attention to here is the difference in the pump-probe traces measured at the same pump and probe photon energies and at different pump and probe photon energies. As mentioned in the Introduction, an additional background contribution appears in the pump-probe traces when probing at the same photon energies as pumping [Figs. 1(c) and 3, 4]. In the corresponding TA spectra, these two types of pump-probe traces correspond to the narrowband VB-AB contribution and the broadband CB-AB contribution, respectively [Figs. 1(b) and 2(a) and (b)].

Figures 3 and 4 show that the VB-AB signal weakens with decreasing film thickness, indicating that it is more likely associated with the ultraslow relaxation process in the bulk of the film. The hole effect also decreases with respect to the electron effect for thinner films. This behavior is confirmed by the ratio of the intensities of the VB-AB contribution to the CB-AB contribution, which sharply decreases with decreasing film thickness [Fig. 5(a)]. On the contrary, the contribution of the inverse bremsstrahlung type FCA compared to the CB-AB contribution increases with decreasing film thickness and tends to saturation [Fig. 5(b)], thus confirming that it is associated with the Dirac SS2.[26]

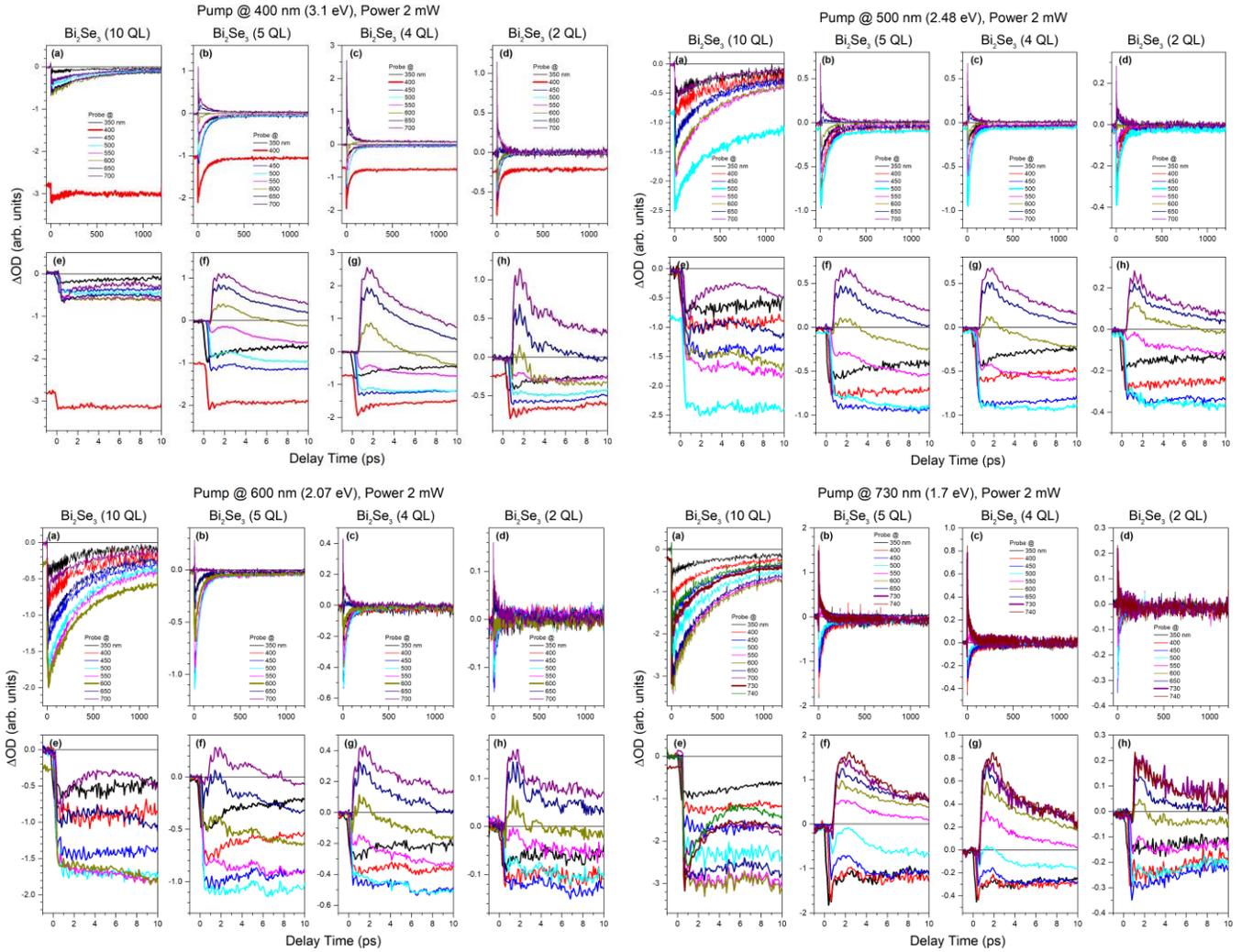

FIG. 4. The same as shown in Fig. 3, but for pumping at 400, 500, 600 and 730 nm (photon energies of 3.1, 2.48, 2.06, and 1.7 eV, respectively) as indicated.

It is noteworthy that the intensity ratio of VB-AB to CB-AB increases more sharply with decreasing pump wavelength (i.e., with increasing pump photon energy) compared to the intensity ratio of FCA to CB-AB [Fig. 5(c) and (d)]. This behavior indicates that the pump photon energy is the main parameter determining the efficiency of excitation of long-lived holes in the bulk states and FCA in the Dirac SS2. However, although both processes become much more efficient at shorter pump wavelengths, their efficiency increases at different rates. As a result, a sharp change in the pumping efficiency of holes indicates that it is not associated with a change in the density of states (absorption coefficient) in the bulk of the film, but rather indicates a nonlinear optical effect. On the contrary, the efficiency of FCA in the Dirac SS2 is most likely controlled by a change in the density of states in the bulk and the subsequent redistribution of electrons towards the Dirac SS2 with decreasing film thickness.[26] In addition, the intensity of the FCA contribution increases almost quadratically with increasing probe wavelength [Fig. 5(e) and (f)], which is in good agreement with the theoretical predictions for the inverse bremsstrahlung type FCA.[30]

The most important observation is the slowly rising component in the pump-probe traces measured at the same pump and probe photon energies. This component accompanies the background signal and is observed mainly for pumping at 340 and 400 nm (photon energies ~3.65 and ~3.1 eV, respectively) [Fig. 6(a)-(e)]. However, since a weak background signal still appears at longer pump wavelengths (Figs. 3 and 4), we argue that the relaxation dynamics measured at all pump wavelengths differ only quantitatively, and the slowly rising component for longer pump wavelengths is too weak to be recognized. Moreover, this tendency for a slow rise and ultraslow decay of the pump-probe signal gradually disappears with decreasing film thickness for all applied pump wavelengths [Figs. 3, 4 and 6(f)–(k)], which also confirms its generality.

We associate this unusual behavior observed at the same pump and probe photon energies with multiphoton photoemission of massless Dirac fermions from the Dirac SS and bound valence electrons from the bulk states. The photoemitted electrons are expected to be trapped at the $Bi_2Se_3/MgF_2$ interface. The role of another interface ($Bi_2Se_3/Al_2O_3$) can be ignored, since the band gap for $\alpha$-$Al_2O_3$ (9 eV)[61] is much larger than for the $MgF_2$ film 10 nm thick (6 eV).[62] The resulting loss of electrons in the film makes it impossible for the remaining holes to recombine and creates their long-lived excess, which is detected as a slowly rising and ultraslowly decaying response (VB-AB contribution). In contrast, one-photon excited and multiphoton excited electrons that were not photoemitted from $Bi_2Se_3$ and trapped at the $Bi_2Se_3/MgF_2$ interface are responsible for the CB-AB contribution. The latter response decays on a much shorter time scale of a few hundred ps and is superimposed on the VB-AB response [Figs. 1(c), 3 and 4]. Nevertheless, the CB-AB response can be observed separately when pumping and probing at different wavelengths. To characterize this ultraslow hole dynamics, we fit the measured pump-probe traces using the appropriate rising and decaying components. The fitting results and corresponding fitting parameters are shown in Figure 7. The criterion for fitting with a long decay time component was to achieve the same background signal intensity as before the action of the pump pulse.

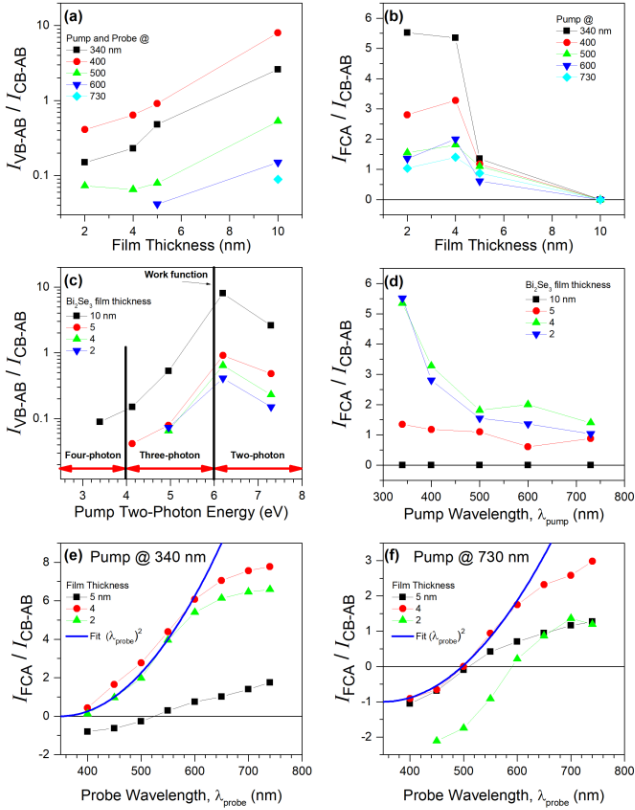

FIG. 5. (a) and (b) Intensity ratios of VB-AB to CB-AB and FCA to CB-AB contributions depending on the $Bi_2Se_3$ film thickness, respectively. The pump and probe wavelengths in (a) were the same, as indicated by the corresponding colors. The pump wavelengths in (b) are indicated by the corresponding colors. The probe wavelength in (b) for the CB-AB contribution was 350 nm, while for the FCA contribution, it was 700 nm for all pump wavelengths except for the 730 nm pump for which 740 nm probe was used. (c) and (d) Pump two-photon energy and pump wavelength dependences of the same intensity ratios shown in (a) and (b), respectively, for different $Bi_2Se_3$ film thicknesses, as indicated by the corresponding colors. In (c) the work function for $Bi_2Se_3$ (5.97 eV) and energy ranges for two-photon, three-photon, and four-photon photoemission are shown. (e) and (f) Probe wavelength dependencies of the intensity ration of the FCA and CB-AB contributions measured as a function of the $Bi_2Se_3$ film thickness for pump wavelengths of 340 and 730 nm, respectively. The probe wavelength for the CB-AB contribution was fixed at 350 nm. An example of a quadratic fit is shown.

Overall, the discussed trapping dynamics is very similar to that occurring at the $Si/SiO_2$ interface under multiphoton (three-photon) pumping in Si, which also provides ultraslow electron dynamics, but on a much longer time scale of several hundred seconds.[52,53] However, in the latter case, the ultraslow dynamics is associated with the transfer of electrons into an ultrathin (less than 10 nm) $SiO_2$ layer at a rate that depends on the thickness of the oxide and the oxygen pressure in the ambient.[52] This behavior contrasts sharply with what happens for the 3D TI $Bi_2Se_3$ in the presence of the $Bi_2Se_3/MgF_2$ interface, where the $MgF_2$ layer is too thick for electron transfer and therefore electron trapping is limited only by the Bi$_2$Se$_3$/MgF$_2$ interface. For this reason, the ultraslow dynamics of holes observed here for the 3D TI Bi$_2$Se$_3$ due to the presence of the Bi$_2$Se$_3$/MgF$_2$ interface is much faster than at the Si/SiO$_2$ interface. In addition, the ultraslow dynamics at the Si/SiO$_2$ interface is characterized by trapping and detrapping rates, which are of the same order of magnitude. This behavior again contrasts with what we observed here for the 3D TI Bi$_2$Se$_3$ in the presence of the Bi$_2$Se$_3$/MgF$_2$ interface. In particular, the slowly rising component has a rise time constant of ~600 ps, while the ultraslowly decaying component, which characterizes the electron detrapping rate, has a decay time constant of about 200 μs. This behavior indicates that the slowly rising component is not related to the electron trapping rate at the Bi$_2$Se$_3$/MgF$_2$ interface and has a different nature.

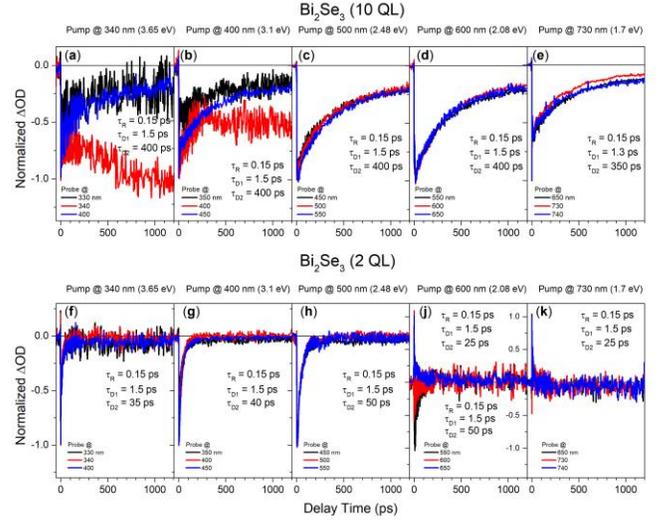

FIG. 6. (a)-(e) and (f)-(k) Normalized pump-probe traces of the 3D TI $Bi_2Se_3$ (10 nm thick film) and the 2D TI $Bi_2Se_3$ (2 nm thick film), respectively, which were measured with pumping at different wavelengths as indicated and probing at three different wavelengths, including the pumping wavelength, as well as slightly shorter and slightly longer than it. The corresponding rise-time and decay-time constants are listed on each plot.

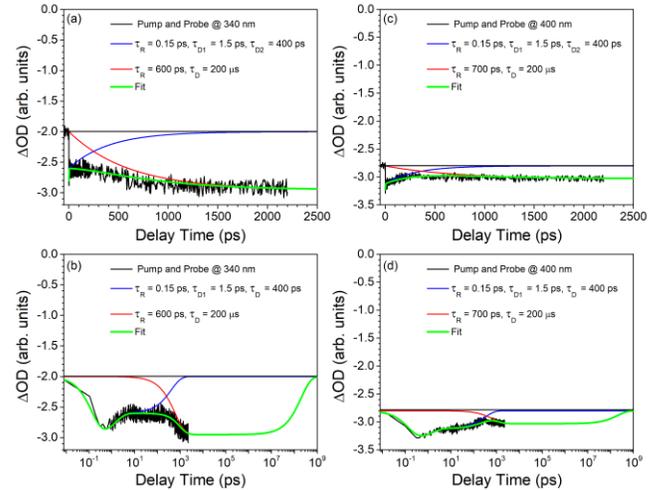

FIG. 7. (a), (b) and (c), (d) Fitting of pump-probe traces of the 3D TI $Bi_2Se_3$ (10 nm thick film) measured at pump and probe wavelengths of 340 nm (3.65 eV) and 400 nm (3.1 eV), respectively. (b) and (d) The same traces shown in (a) and (c), but on a semi-logarithmic scale. Fitting components and the rise-time and decay-time constants are indicated by the corresponding colors.

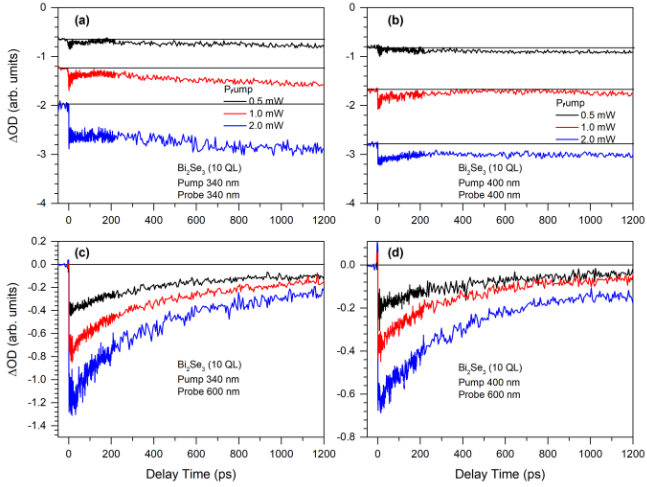

FIG. 8. Pump-probe traces of the 3D TI Bi$_2$Se$_3$ (10 nm thick film) measured at different pump powers, as indicated by the corresponding colors, and at the same pump and probe photon energies [(a) and (b)] and at different pump and probe photon energies [(c) and (d)], as indicated.

The discussed two types of pump-probe traces and the slowly rising component can also be clearly recognized by their pump power dependencies (Fig. 8). Correspondingly, the intensity of the pump-probe signal, measured at different pump and probe photon energies (CB-AB contribution), increases sublinearly with increasing pump power [Fig. 8(c) and (d)]. In addition to this traditional trend,[26] the VB-AB signal appears exclusively when pumping and probing at the same photon energy, and its intensity also increases sublinearly with increasing pump power [Fig. 8(a) and (b)]. Both tendencies are in good agreement with that observed when using a laser with a pulse repetition rate of 80 MHz.[11] However, for the pump-probe traces measured at the same photon energies, the slowly rising component appears, which significantly modifies the usual decay trend, forming a characteristic inflection point [Fig. 8(a) and (b)]. The presence of an inflection point clearly indicates the competition of two relaxation dynamics of the opposite sign. Moreover, the slowly rising component manifests itself separately as the minimum pump power is applied [Fig. 8(a) and (b)]. This behavior indicates that the decay trend begins to dominate over the slowly rising trend only with increasing pump power. Accordingly, at lower pump powers, the main role is played by multiphoton (presumably two-photon for 340 and 400 nm pump wavelengths) photoemission and subsequent electron trapping at the Bi$_2$Se$_3$/MgF$_2$ interface, while electron-hole recombination begins to dominate only as the density of photoexcited electrons in the system increases with increasing pump power.

These complex trends of photoexcited carriers can be understood by considering the band structure of the 3D TI Bi$_2$Se$_3$ [Fig. 2(c)].[23,35] Specifically, optical pumping is expected to occur in three principal ways: (i) from the full-filled states of the upper Dirac cone of the Dirac SS1 below the Fermi energy ($E_F$); (ii) from the full-filled states at the VB top; (iii) from the half-filled states associated with the VB SOC split components (SOCS1 and SOCS2) arising due to strong SOC and inversion symmetry breaking near the surface.[23] However, since the high-energy edge of the measured broadband CB-AB contribution significantly exceeds all applied pump photon energies, we consider the multiphoton pumping of massless Dirac fermions, initially located in the upper Dirac cone of Dirac SS1 below $E_F$, towards the higher energy Dirac SS (SS2, SS3 and SS4) as the main nonlinear mechanism leading to their excitation. Here we especially emphasize that the multiphoton pumping of massless Dirac fermions is expected to occur not only for 1.7 eV photons, as previously reported for two-photon pumping,[26,28] but also for all pump photon energies used in this study.

Consequently, the pump photon energy completely determines the order of nonlinearity of optical processes leading to the excitation of massless Dirac fermions and their photoemission, since the condition must be met under which the total energy of multiphoton excitation exceeds the work function of Bi$_2$Se$_3$ (5.97 eV) [Fig. 5(c)].[28,35] Because the higher order nonlinearity (higher order multiphoton process) is required for longer pump wavelengths, the efficiency of multiphoton photoemission (and the intensity ratio of VB-AB to CB-AB) drops sharply with decreasing pump photon energy [Figs. 3, 4, and 5(c)]. The high efficiency of multiphoton pumping of massless Dirac fermions is associated with their giant nonlinearity in Dirac SS.[28,63,64] Moreover, the multiphoton pumping of massless Dirac fermions is expected to occur resonantly due to the continuum nature of the Dirac SS extending over a wide energy range [Fig. 2(c)].[27,35] Consequently, this behavior might include cascade one- and two-photon absorption, or cascade one- and three-photon absorption, or direct two-photon, three-photon or even four-photon absorption depending on the pump photon energy and resonance conditions for the higher energy Dirac SS in the band structure of Bi$_2$Se$_3$ [Figs. 2(c) and 5(c)].

However, the multiphoton pumping of massless Dirac fermions occurs simultaneously with the pumping of the bound valence electrons from the VB top and VB SOC split states (SOCS1 and SOCS2) of the film bulk.[23] In contrast to massless Dirac fermions, the pumping of bound valence electrons can be either one-photon or multiphoton, depending on how resonantly the pump photon energy coincides with the density of states in various CB (VB) of Bi$_2$Se$_3$ [Fig. 2(c)]. For this reason, the resonance feature observed in TA spectra at 400 nm pumping [Fig. 2(a)] indicates that in this case the resonance conditions for multiphoton (presumably two-photon) photoemission of bound valence electrons are much better compared to other pump wavelengths. A special case is the photoemission of bound valence electrons from the VB SOCS2 state, which for the lowest order nonlinearity (two-photon process) is possible only with pumping at 340 nm [Fig. 2(c)]. Further relaxation of electrons not subjected to multiphoton photoemission and subsequent trapping at the Bi$_2$Se$_3$/MgF$_2$ interface determines the broadband CB-AB contribution expanding from their initial excitation energy towards lower energy bulk states and Dirac SS, which are successively filled upon relaxation. This behavior means that if the multiphoton pumping energy is insufficient for multiphoton photoemission or the resonance conditions for multiphoton pumping are poorly met, then photoexcited carriers exhibit ordinary electron–hole recombination on a time scale of several hundred ps, and ultraslow hole dynamics are not observed.

Here we emphasize that the multiphoton transitions in the film bulk occur from all available full-filled and half-filled VB states to all available CB states according to their density of states, although Figure 2(c) only shows the principal one-photon transitions discussed. Despite such a variety of photoexcitations, the photoexcited holes finally accumulate at the edge of the VB and the lower Dirac cone of the Dirac SS1, blocking out the corresponding probing optical transitions originating from these states. The resulting narrowband VB-AB contribution images hence this relaxation process of photoexcited holes in the bulk of the film [Fig. 1(b), Fig. 2(a) and (b)]. However, the hole dynamics also includes the intervalley relaxation between the SOCS components [Fig. 2(c)],[23] similarly to how it occurs in transition metal dichalcogenides.[20-22] Since the latter process continuously feeds states at the VB edge and requires a spin-flip, the VB-AB contribution exhibits a slowly rising component with a rise-time constant of 600 ps, which therefore has nothing to do with the electron trapping rate. As regards the electron trapping rate, it is expected that it will characterize the dynamics on a subpicosecond time scale, similar to what observed in Mn-doped $Bi_2Se_3$ films.[65]

The main question is why the ultraslow hole dynamics manifests itself exclusively in a narrow spectral range near the edge of the VB and the lower Dirac cone of the Dirac SS1. The answer is dictated by experimental conditions. Specifically, the transient absorption coefficient can be expressed as $\alpha = \alpha_0(1 - f_e - f_h)$,[4] where $f_e$ and $f_h$ are the Fermi-Dirac occupancy factors for electrons and holes, respectively, and $\alpha_0$ is the absorption coefficient with no pump applied. This presentation implies that the transient response associated with AB in a film of thickness $d$ can be presented as $\Delta OD = d\Delta\alpha = -d\alpha_0(f_e + f_h)$,[30] thus being negative once the pump-excited electrons (holes) fill up the band states in the CB (VB) in accordance with their occupancy factors (CB-AB and VB-AB, respectively). Since photoexcited holes are accumulated at the edge of the VB and the lower Dirac cone of the Dirac SS1 [Fig. 2(c)], and the energy of optical transitions also appears as measured from the top of the VB, both occupancy factors affect the relaxation dynamics only when pumping and probing at the same photon energies. The corresponding intensity of AB response is expected to be the sum of the VB-AB and CB-AB contributions [Fig. 1(b)]. On the contrary, if the energies of the pump and probe photons are different, the electronic dynamics prevails and only the CB-AB contribution appears. Consequently, the photoexcited holes at the edge of the VB and the lower Dirac cone of the Dirac SS1 make a dominant contribution to the narrowband feature in TA spectra associated with the VB-AB signal. The reason is that the half-filled SOCS states remain inactive for the linearly polarized probe light used in this study. This behavior arises because the linearly polarized light can be considered as a superposition of two circularly polarized waves propagating in the same direction but the opposite sense of rotation. As a result, a linearly polarized probe pulse in the half-filled SOCS states will simultaneously experience two transient effects (bleaching and absorption) that cancel each other out.

The ultraslow hole dynamics is sharply suppressed with decreasing film thickness for the 2D TI $Bi_2Se_3$ (film thicknesses 5, 4, and 2 QL) [Fig. 5(a)]. This behavior suggests that the multiphoton photoemission of Dirac fermions becomes inefficient due to the loss of resonance conditions caused by the gap opening at the Dirac SS nodes for films thinner than 6 QL.[46] For the same reason, massless Dirac fermions gain mass due to time-reversal symmetry breaking.[46-51] Such an abrupt change in the pumping efficiency for massive Dirac fermions will affect the relaxation dynamics, even if multiphoton photoemission of bound valence electrons still takes place. Accordingly, non-photoemitted massive Dirac fermions in the ultrathin $Bi_2Se_3$ film will accumulate in the Dirac SS2 and recombine with photoexcited holes accumulated in the Dirac SS1,[26] thereby eliminating their ultraslow dynamics.

In addition, the suppression of multiphoton photoemission of massive Dirac fermions for films thinner than 6 QL also leads to an increase in their density. This behavior manifests itself in an increase in the intensity of the FCA contribution as the pump wavelength decreases [Fig. 5(d)]. Since this dependence on the pump wavelength is not as sharp as for the VB-AB contribution [Fig. 5(c)], we attribute it to a higher density of states in the CB for higher energies. Consequently, further redistribution of photoexcited electrons towards the Dirac SS2 with decreasing film thickness leads to an increase in the density of massive Dirac fermions observed. Thus, the massive Dirac fermions predominantly determines the relaxation of photoexcited carriers for both the 2D topologically nontrivial and 2D topologically trivial insulator phases.

## Conclusions

In summary, using UV-Vis TA spectroscopy in the range of 1.65 – 3.8 eV we have shown that when pumping with different wavelengths of 340, 400, 500, 600, and 730 nm (the corresponding photon energies of 3.65, 3.1, 2.48, 2.07, and 1.7 eV, respectively), the TA spectra of the 2D and 3D TI $Bi_2Se_3$ coated with a 10 nm thick $MgF_2$ layer cover the entire visible and part of the UV region. This observation suggests that multiphoton excitation accompany the usual one-photon excitation. Correspondingly, for the 3D TI $Bi_2Se_3$, we observed ultraslow hole dynamics associated with multiphoton resonant photoemission of massless Dirac fermions and bound valence electrons and their subsequent trapping at the $Bi_2Se_3/MgF_2$ interface. The ultraslow hole dynamics is hence due to the electron deficit in the film, which makes it impossible for the remaining holes to recombine and creates their long-lived excess. This tendency is detected at a specific probing wavelength as a slowly rising and ultraslowly decaying response. The corresponding rise- and decay-time constants equal to 600 ps and 200 µs characterize a large SOC splitting at the VB maximum and the electron detrapping rate, respectively. This behavior sharply decreases with decreasing pump photon energy, indicating that multiphoton pumping occurs for all used pump wavelengths but involves different orders of optical nonlinearity. We attributed the high efficiency of multiphoton pumping to the giant nonlinearity of massless Dirac fermions in the Dirac SS and resonance-enhanced transitions from the bulk states. On the contrary, if the energy of multiphoton pumping is insufficient for multiphoton photoemission or the resonance conditions for multiphoton pumping are poorly met, then photoexcited carriers demonstrate the usual electron-hole recombination on a time scale of several hundred ps, and ultraslow hole dynamics are not observed.

The observed ultraslow hole dynamics is very similar to that occurring at the $Si/SiO_2$ interface under multiphoton (three-photon) pumping in Si, which also provides ultraslow electron dynamics, but on a much longer time scale of several hundred seconds. The dynamics of long-lived holes is gradually suppressed with decreasing film thickness for the 2D TI $Bi_2Se_3$ due to the loss of resonance conditions caused by the gap opening at the Dirac SS nodes. This behavior indicates that the massive Dirac fermions predominantly determine the relaxation of photoexcited carriers for both the 2D topologically nontrivial and 2D topologically trivial insulator phases.

Finally, we conclude that the 3D TI $Bi_2Se_3$ coated with a thin $MgF_2$ layer can be applied to create new thin-film quantum materials that are promising for modern optoelectronics and optospintronics.

**Acknowledgements**
This work was supported by the National Key Research and Development Program of China administrated by the Ministry of Science and Technology of China (Grant No. 2016YFB0401702), the Shenzhen Peacock Team Project (Grant No. KQTD2016030111203005), and the Shenzhen Key Laboratory for Advanced Quantum Dot Displays and Lighting (Grant No. ZDSYS201707281632549). The authors acknowledge J. Li for help with the laser system operation and S. Babakiray for growing the $Bi_2Se_3$ samples by MBE (under supervision of D. Lederman) using the West Virginia University Facilities.